\documentclass[twocolumn,showpacs,preprintnumbers,amsmath,amssymb]{revtex4}

\usepackage{graphicx}
\usepackage{dcolumn}
\usepackage{bm}
\usepackage{enumerate}

\begin{document}

\preprint{To be published in Phys. Rev. B}

\title{Oxygen-related traps in pentacene thin films:\\ Energetic position and implications for transistor performance}

\author{Wolfgang L. Kalb}%
 \email{kalb@phys.ethz.ch}
\author{Kurt Mattenberger}
\author{Bertram Batlogg}
\affiliation{%
Laboratory for Solid State Physics, ETH Zurich, 8093 Zurich,
Switzerland
}%

\date{\today}

\begin{abstract}
We studied the influence of oxygen on the electronic trap states in
a pentacene thin film. This was done by carrying out gated
four-terminal measurements on thin-film transistors as a function of
temperature and without ever exposing the samples to ambient air.
Photooxidation of pentacene is shown to lead to a peak of trap
states centered at 0.28\,eV from the mobility edge, with trap
densities of the order of 10$^{18}$\,cm$^{-3}$. These trap states
need to be occupied at first and cause a reduction in the number of
free carriers, i.e. a consistent shift of the density of free holes
as a function of gate voltage. Moreover, the exposure to oxygen
reduces the mobility of the charge carriers above the mobility edge.
We correlate the change of these transport parameters with the
change of the essential device parameters, i.e. subthreshold
performance and effective field-effect mobility. This study supports
the assumption of a mobility edge for charge transport, and
contributes to a detailed understanding of an important degradation
mechanism of organic field-effect transistors. Deep traps in an
organic field-effect transistor reduce the effective field-effect
mobility by reducing the number of free carriers and their mobility
above the mobility edge.
\end{abstract}

\maketitle

\section{\label{sec:level1} Introduction}

Organic semiconductors are among the most promising candidates for
the future's flexible and low-cost electronics. Apart from the
processability from solution or by thermal evaporation, a huge
potential lies in tailoring the properties of the organic
semiconductor by means of synthetic organic chemistry.

Charge carrier mobilities in organic field-effect transistors are
comparable to the mobilities in hydrogenated amorphous silicon
transistors and are adequate for many applications.\cite{LinYY19972,
GundlachDJ2008} Critical issues, however, are electrical stability
and environmental stability. The electrical stability of both n- and
p-type organic semiconductor transistors has recently been shown to
be very high if suitable gate dielectrics are
used.\cite{KalbWL2007,ZhangXH2007} The environmental stability of
the organic semiconductor, thus, is an urgent issue to be
addressed.\cite{deLeeuwDM1997, AnthopoulosTD2007} Studies of the
degradation of organic field-effect transistors are rare and
indicate that, in the case of p-type organic semiconductors,
atmospheric oxygen or ozone is a major cause.\cite{PannemannC2004,
MaliakalA2004, DeAngelisF2006, ChabinycML2007, KlaukH2007} It is
crucial to understand in detail the way in which an oxidation of the
organic semiconductor impedes the charge transport and thus degrades
the transistor characteristics.

The charge transport in crystalline organic semiconductors such as
pentacene may be described by assuming a mobility edge which
separates extended and localized states. The charge carriers are
transported in the extended states above the mobility edge, but are
trapped by and thermally released from localized trap states below
the mobility edge.\cite{AndersonPW1972, WartaW1985, HorowitzG1995,
PernstichKP2008} The mobility edge may be identified with the
valence or conduction band edge. However, the parameters dominating
charge transport are the trap densities as a function of energy
relative to the mobility edge, the number of delocalized charge
carriers above the mobility edge and the mobility of the latter
charge.
In highly disordered organic semiconductors, a description of the
charge transport by variable range hopping may be more appropriate.
Importantly, this situation can be described by trap-controlled
transport in a transport level with a distribution of localized
states below the transport level and is thus very similar to the
mobility edge picture.\cite{ArkhipovVI2003}

Organic field-effect transistors are excellent tools to study the
charge transport in organic semiconductors since the position of the
quasi-Fermi level at the dielectric-semiconductor interface can be
fine-tuned by applying a gate voltage.\cite{HorowitzG1995,
SchauerF1999, LangDV2004, SalleoA2004, CalhounMF2007, KawasakiN2007,
PernstichKP2008}

Here we report on an extended study of pentacene thin-film
transistors. Organic field-effect transistors were characterized by
temperature dependent measurements without ever exposing the samples
to ambient air and after controlled exposure to oxygen and light.
The effect of oxygen can only be clarified with pristine samples.
Moreover, since the field-effect conductivity depends
(approximately) exponentially on temperature, temperature is a very
sensitive parameter.

Another distinct feature of our study are the gated four-terminal
measurements instead of the commonly employed gated two-terminal
measurements. This approach has previously been used to estimate
basic parameters, such as the effective field-effect mobility and
the contact resistance.\cite{TakeyaJ2003, PesaventoPV2004}. The
contact resistance can significantly affect the field-effect
mobility. Consequently, contact effects may also lead to errors e.g.
in the trap densities as extracted from transistor characteristics.

In order to assess the fundamental transport parameters we have
developed a scheme for organic field-effect transistors that is easy
to use. The approach readily reveals all the key parameters with
high accuracy in a straightforward and unambiguous fashion. The
scheme is based on a method developed by Gr\"unewald et al. for
amorphous silicon field-effect transistors.\cite{GrunewaldM1980}
Instead of estimating the interface potential from the transistor
characteristic measured at a single temperature as in the original
scheme, we extract the interface potential from the
temperature-dependence of the field-effect conductivity.

\section{Experimental}

The system used in this study allows for the fabrication and
characterization of organic thin-film transistors without breaking
the high vacuum of order 10$^{-8}$\,mbar between the fabrication and
measurement steps. It consists of a prober station connected to an
evaporation chamber. The prober station contains a cryostat for
temperature-dependent measurements. A schematic drawing of the
system can be found in Ref.~\onlinecite{KalbWL20072}.

\begin{figure}
\includegraphics[width=0.90\linewidth]{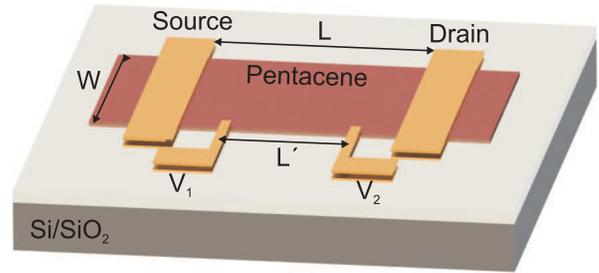}
\caption{\label{device} (Color online) The transistors for the gated
four-terminal measurements consisted of a well-defined stripe of
pentacene with a width of $W=1000$\,$\mu$m and gold electrodes. The
distance between source and drain was $L=450$\,$\mu$m. The voltage
sensing electrodes were used to measure the potentials $V_{1}$ and
$V_{2}$ with respect to the grounded source and were separated by
$L'=300$\,$\mu$m.}
\end{figure}

Heavily doped Si wafers with a 260\,nm thick SiO$_{2}$ layer were
cut, cleaned with hot solvents, fixed on a sample holder and were
then introduced into the cryo-pumped evaporation chamber of the
device fabrication and characterization system. After approximately
24\,h, two times sublimation purified pentacene was evaporated onto
the samples through a shadow mask at a base pressure of the order of
10$^{-8}$\,mbar. The substrates were kept at room temperature during
the evaporation and the final film thickness was 50\,nm. The sample
holder with the samples was then placed on a shadow mask for the
gold evaporation without breaking the high vacuum and gold
electrodes were evaporated onto the pentacene. The completed
thin-film transistors had a channel length of $L=450$\,$\mu$m and a
channel width of $W=1000$\,$\mu$m. Moreover, two voltage sensing
electrodes were situated at $(1/6)L$ and $(5/6)L$ and had little
overlap with the ``masked'' pentacene film, as schematically shown
in Fig.~\ref{device}. The alignment was achieved by means of a high
precision mask positioning mechanism. The sample holder was then
transferred to and clamped on the cryostat in the turbo-pumped
prober station of the device fabrication and characterization system
without breaking the high vacuum (10$^{-8}$\,mbar).

A previous study revealed that the device performance improves with
time when pentacene thin-film transistors are kept in high
vacuum.\cite{KalbWL20072} Therefore, the devices were kept in the
prober station at a pressure of $\approx2\times10^{-8}$\,mbar for
approximately three weeks before starting the study. After that
time, the device characteristics were found to be stable on the
timescale of several days.

The prober station is equipped with five micro-probers. The prober
arms are connected to the cryostat with thick copper braids and are
thus cooled when the cryostat and the sample holder with the samples
is cooled down. For the temperature-dependent gated four-terminal
measurements, the source, the drain and the voltage sensing
electrodes were contacted with thin gold wires attached to four of
the micro-probers. By means of an electrical feedthrough to the
cryostat, a gate bias could be applied. In order to measure the
temperature on the surface of the samples, an AuFe/Chromel
thermocouple was attached on the 5th micro-prober and was carefully
pressed against the surface of the sample at each temperature.

The electrical measurements were carried out with an HP 4155A
semiconductor parameter analyzer. Transfer characteristics were
measured with a drain voltage of $V_{d}=-2$\,V in steps of 0.2\,V
with an integration time of 20\,ms and a delay time of 100\,ms. For
each gate voltage $V_{g}$, the drain current $I_{d}$ and the
potentials $V_{1}$ and $V_{2}$ between the grounded source electrode
and the respective voltage sensing electrode were measured
(Fig.~\ref{device}). All electrical measurements were done in the
dark.

The devices were exposed to oxygen by introducing 1\,bar of oxygen
(purity $\geq99.9999$\,vol$\%$) into the prober station through a
leak valve. The pressure of the oxygen in the prober station was
measured with a mechanical pressure gauge. In addition, the samples
were also exposed to a combination of 1\,bar of oxygen and white
light from a cold light source (color temperature 3200\,K).

\section{Charge transport parameters}

Our approach consists of measuring the transfer characteristics at a
low drain voltage ($V_{d}=-2$\,V). This may be understood as staying
as close as possible to the ``unperturbed'' situation where charge
is accumulated by a gate voltage in a metal-insulator-semiconductor
(MIS) structure but no drain voltage is applied.\cite{HorowitzG2004}
We begin by specifying the basic parameters to be extracted from the
transfer characteristics.

\subsection{Field-effect conductivity, field-effect mobility and contact resistance}

As described in Ref.~\onlinecite{KalbWL20072} in more detail, the
field-effect conductivity $\sigma$ is given by
\begin{equation} \label{sigma2P}
\sigma(V_{g})=\frac{L}{W}\frac{I_{d}}{V_{d}}.
\end{equation}
and is free from contact effects when calculated with
\begin{equation}\label{sigma4P}
\sigma(V_{g})=\frac{L'}{W}\frac{I_{d}}{(V_{1}-V_{2})}.
\end{equation}
The field-effect mobility $\mu_{eff}$ in the linear regime is
generally estimated with
\begin{equation} \label{mu2P}
\mu_{eff}(V_{g})=\frac{L}{WV_{d}C_{i}}\bigg(\frac{\partial
I_{d}}{\partial V_{g}}\bigg)_{V_{d}}.
\end{equation}
This mobility is not influenced by the contact resistance if
calculated with
\begin{equation}\label{mu4P}
\mu_{eff}(V_{g})=\frac{L'}{W(V_{1}-V_{2})C_{i}}\left(\frac{\partial
I_{d}}{\partial V_{g}}\right)_{V_{d}}.
\end{equation}
We use the terms ``two-terminal conductivity'' and ``four-terminal
conductivity'' for the expressions defined respectively in
Eq.~\ref{sigma2P} and Eq.~\ref{sigma4P}. Eq.~\ref{mu2P} is the
``two-terminal mobility'' and Eq.~\ref{mu4P} is the ``four-terminal
mobility''. $\mu_{eff}$, as calculated with Eq.~\ref{mu2P} or
Eq.~\ref{mu4P}, is an effective mobility. For a p-type semiconductor
such as pentacene, it is a rough estimate of the ratio of the free
surface hole density $P_{free}$ to the total surface hole density
$P_{total}$ multiplied with the mobility $\mu_{0}$ of the holes in
the valence band, i.e.
\begin{equation} \label{allgmu}
\mu_{eff}\approx\frac{P_{free}}{P_{total}}\mu_{0}.
\end{equation}
Assuming a linear voltage drop all along the transistor channel the
total contact resistance can be approximated as
\begin{equation}
R_{contact}(V_{g})=\frac{V_{d}-(V_{1}-V_{2})L/L'}{I_{d}}
\end{equation}
and should be compared to the channel resistance of the device as
given by
\begin{equation}
R_{channel}(V_{g})=\frac{(V_{1}-V_{2})L/L'}{I_{d}}.
\end{equation}

In order to gain a deeper insight into the physics of the organic
semiconductor in a field-effect transistor, a more sophisticated
description is developed in the following.

\subsection{Spectral density of trap states and free hole density} \label{demo1}

The transistor characteristics critically depend on trap states. We
treat the polycrystalline pentacene layer as uniform and assume that
trap states on the surface of the gate dielectric only contribute to
a non-zero flatband voltage. Consequently, the trap densities to be
determined are an average over in-grain and grain boundary regions
and may also be influenced, to some extend, by trap states on the
surface of the gate dielectric.

\begin{figure}
\includegraphics[width=0.60\linewidth]{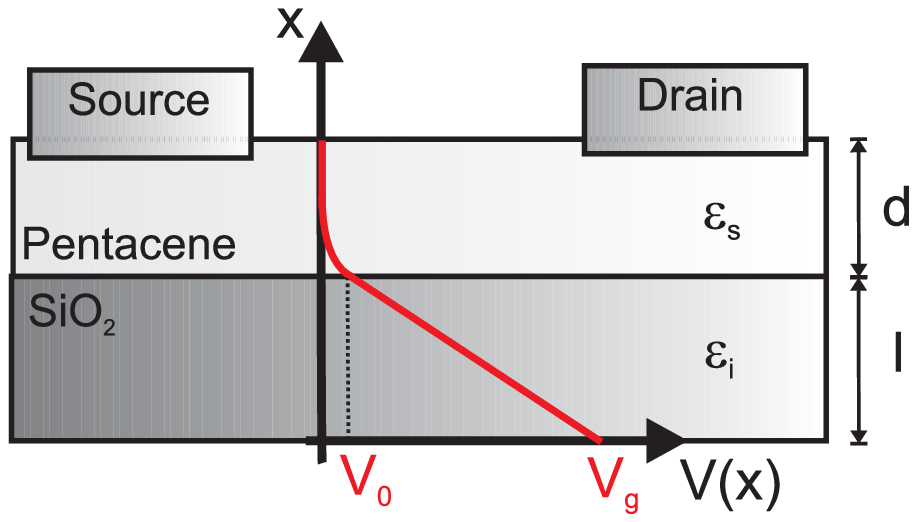}
\caption{\label{transistor} (Color online) Potential drop across the
gate insulator (thickness l, dielectric constant $\epsilon_{i}$) and
the pentacene thin film (thickness d, dielectric constant
$\epsilon_{s}$). Most of the gate voltage drops across the gate
dielectric. At the insulator-semiconductor interface the potential
is $V_{0}$.}
\end{figure}

In the following, we outline how a straightforward conversion of
Poisson's equation along with boundary conditions eventually leads
to the key equations. We assume that the electrical potential $V(x)$
at the surface of the pentacene film of thickness $d$ vanishes under
all biasing conditions, i.e.
\begin{equation}
V(x=d)=0.
\end{equation}
The electric field $F$ at this position is also assumed to drop to
zero:
\begin{equation}
F=-\frac{d}{dx}V(x=d)=0.
\end{equation}
This is reasonable as long as the pentacene film is thicker than the
decay length of the potential.\cite{GrunewaldM1980} The situation is
depicted in Fig.~\ref{transistor}. The dielectric strength at the
insulator-semiconductor interface must be continues, i.e. for a zero
flatband voltage
\begin{equation} \label{dielectricstrength}
\epsilon_{i}\frac{V_{g}-V_{0}}{l}=-\epsilon_{s}\frac{d}{dx}V(x=0).
\end{equation}
$l$ is the thickness of the gate dielectric, $\epsilon_{i}$ and
$\epsilon_{s}$ are the dielectric constants of the gate insulator
and the semiconductor and $V(x=0)=V_{0}$ is the interface potential.
As detailed in Ref.~\onlinecite{GrunewaldM1980}, a conversion of
Poisson's equation with these boundary conditions eventually leads
to an expression for the total hole density $p$ as a function of the
interface potential $V_{0}$:
\begin{equation} \label{holes}
p(V_{0})=\frac{\epsilon_{0}\epsilon_{i}^{2}}{\epsilon_{s}l^{2}e}U_{g}\left(\frac{dV_{0}}{dU_{g}}\right)^{-1}.
\end{equation}
$p$ (lower case) denotes a volume density of holes. The volume
density depends on the distance $x$ from the insulator-semiconductor
interface, i.e. on the electrical potential $V(x)$ in the
semiconductor. The volume density $p$ and the surface hole density
$P$ are related by integrating over the depth of the whole film,
i.e. $P=\int_{0}^{d} p(x)dx$. Eq.~\ref{holes} yields the functional
dependence of the volume density of holes on the potential $V_{0}$.
Moreover,
\begin{equation}
U_{g}=|V_{g}-V_{FB}|
\end{equation}
in Eq.~\ref{holes} is the gate voltage above the flatband voltage
$V_{FB}$. Since the total hole density $p$ can be written as
\begin{equation}
p(V)=\int_{-\infty}^{+\infty}N(E)\left[f(E-eV)-f(E)\right]dE,
\end{equation}
its derivative is given by
\begin{equation} \label{holedensity}
\frac{1}{e}\frac{dp(V)}{dV}=\int_{-\infty}^{+\infty}N(E)\left|\frac{df(E-eV)}{d(E-eV)}\right|dE.
\end{equation}
Eq.~\ref{holedensity} is a convolution of the density of states
function $N(E)$ with the derivative of the Fermi function. We
approximate the Fermi function with a step function according to the
common zero-temperature approximation.\cite{HorowitzG1995,
KalbWL20072, BragaD2008} Its derivative then is a delta function and
we eventually have
\begin{equation} \label{DOS}
\frac{1}{e}\frac{dp(V_{0})}{dV_{0}}\approx N(E_{F}+eV_{0}).
\end{equation}

\begin{figure}
\includegraphics[width=0.55\linewidth]{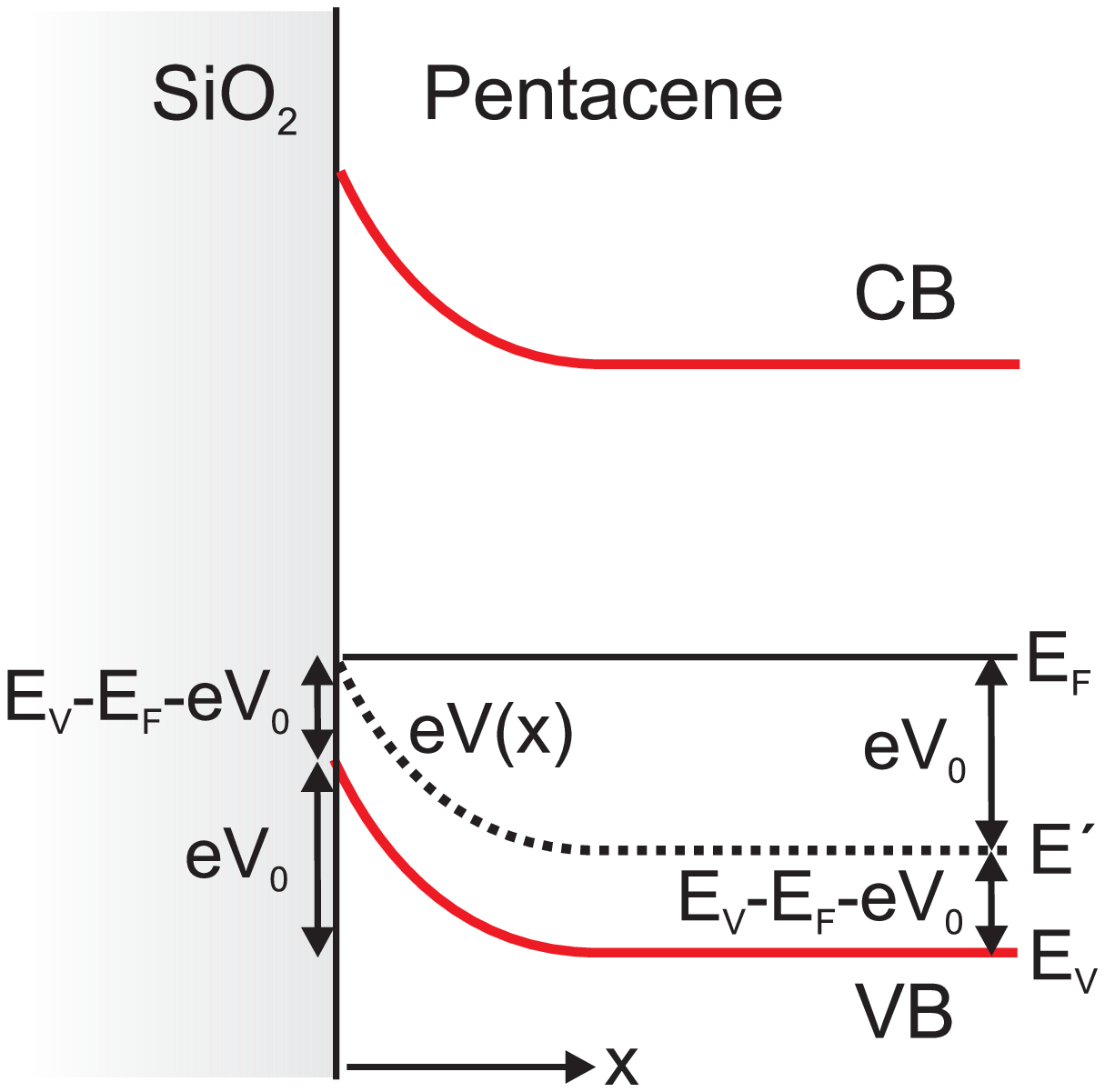}
\caption{\label{bandalt} (Color online) Gate voltage induced band
bending at the insulator-semiconductor interface. At a given gate
voltage, the band shift at the interface is $eV_{0}$. The energy of
trap states with an energy $E'$ is raised at the interface and now
coincides with the Fermi energy $E_{F}$ of the sample. The energy of
these trap states relative to the energy of the mobility edge
$E_{V}$ is $E_{V}-E_{F}-eV_{0}$ and is approximated by the
experimentally available activation energy $E_{a}$ of the
field-effect conductivity.}
\end{figure}

From Eq.~\ref{holes} and Eq.~\ref{DOS} we can see that the interface
potential $V_{0}$ as function of the gate voltage is the key to the
density of states function (DOS). Since the change of the interface
potential and the change of the drain current with gate voltage are
linked, it is possible to extract the interface potential from the
transfer characteristic measured at a single
temperature.\cite{GrunewaldM1980} Once the interface potential is
known, the trap DOS in the mobility gap can be estimated with
Eq.~\ref{holes} and Eq.~\ref{DOS}. Thus, the trap densities can be
plotted as a function of the band shift $eV_{0}$ at the interface,
which is the energy of the traps relative to the Fermi energy of the
sample (Fig.~\ref{bandalt}). In a previous study we have applied
this method to four-terminal conductivity data.\cite{KalbWL20072}
For the present study we have advanced the extraction scheme. As
described in the following, we used gated four-terminal measurements
at various temperatures to estimate the interface potential. We then
used Eq.~\ref{holes} and Eq.~\ref{DOS} to calculate the DOS. The
consistency of the assumption of charge transport above a mobility
edge with the temperature-dependent measurements provides a greater
degree of confidence to any conclusion. Moreover, the latter
approach has the advantage of eventually giving the DOS as a
function of energy relative to the mobility edge.

We now show that the activation energy $E_{a}(V_{g})$ of the
field-effect conductivity as defined by
\begin{equation} \label{sigmaacti}
\sigma(V_{g})=A\exp\left(-\frac{E_{a}}{kT}\right)
\end{equation}
and as determined with Arrhenius plots is related to the band shift
$eV_{0}$ at the insulator-semiconductor interface. Following
Boltzmann's approximation, the field-effect conductivity
\begin{equation} \label{sigmanorm}
\sigma(V_{g})=e\mu_{0}P_{free}
\end{equation}
may be written as
\begin{eqnarray} \label{sigmaboltz}
\sigma(V_{g})=e\mu_{0}\int_{0}^{d}p_{free}dx= \nonumber \\
=e\mu_{0}N_{V}\exp\left(-\frac{E_{V}-E_{F}}{kT}\right)\times\int_{0}^{d}\exp\left(\frac{eV(x)}{kT}\right)dx.
\end{eqnarray}
$N_{V}$ is the effective (volume) density of extended states,
$E_{V}$ is the energetic position of the mobility edge and $E_{F}$
is the Fermi energy. As an example we consider an exponential trap
DOS
\begin{equation}
N(E)=N_{0}\exp\left(\frac{E}{kT_{0}}\right)
\end{equation}
with a characteristic slope of $kT_{0}$.
If we assume that all the gate-induced charge is trapped,
integration of the exponential trap DOS leads to a simple
exponential dependence of the total hole density $p$ on the
potential $V(x)$, which is
\begin{equation} \label{revision1}
p\propto\exp\left(\frac{eV}{kT_{0}}\right).
\end{equation}
This approximation is not expected to lead to serious errors as long
as the majority of the gate-induced charge is
trapped.\cite{SpearWE1972} With Eq.~\ref{revision1} it can be shown
that, except for small values of $V$, also the electric field $F$
perpendicular to the insulator-semiconductor interface exponentially
depends on $V$, as
\begin{equation} \label{revision2}
F\propto\exp\left(\frac{eV}{2kT_{0}}\right).
\end{equation}
Ref.~\onlinecite{HorowitzG1995} shows in detail how the electric
field as in Eq.~\ref{revision2} results in an expression for the
free surface hole density $P_{free}$ in Eq.~\ref{sigmanorm}. This
expression is
\begin{eqnarray} \label{pfree}
P_{free}\approx\frac{2kT\epsilon_{0}\epsilon_{s}}{eC_{i}U_{g}}\frac{l}{2l-1}N_{V}\exp\left(-\frac{E_{V}-E_{F}}{kT}\right)\times\nonumber \\
\times\left[\exp\left(\frac{eV_{0}}{kT}\right)-\exp\left(\frac{eV_{0}}{2lkT}\right)\right]
\end{eqnarray}
where
\begin{equation}
l=\frac{T_{0}}{T}.
\end{equation}
The second exponential term in Eq.~\ref{pfree} can safely be
neglected and so we have
\begin{equation} \label{pfreeapprox}
P_{free}\approx
L_{a}\frac{l}{2l-1}N_{V}\exp\left(-\frac{E_{V}-E_{F}-eV_{0}}{kT}\right),
\end{equation}
with
\begin{equation}
L_{a}=\frac{2kT\epsilon_{0}\epsilon_{s}}{eC_{i}U_{g}}.
\end{equation}
The factor $l/(2l-1)$ is expected to be close to unity and $L_{a}$
may be understood as the effective thickness of the accumulation
layer.\cite{HorowitzG2000}
Small deviations from an exponential trap DOS might be considered
with a variable parameter $l$. However, the variation of $l$ can be
ignored when compared to the exponential term in
Eq.~\ref{pfreeapprox}.
If we compare Eq.~\ref{pfreeapprox} with Eq.~\ref{sigmaacti} and
Eq.~\ref{sigmanorm}, we eventually have
\begin{equation} \label{energy}
E_{a}\approx E_{V}-E_{F}-eV_{0}=E_{V}-E'_{F}.
\end{equation}
The measured activation energy of the field-effect conductivity
$E_{a}(V_{g})$ is approximately equal to the energetic difference
between the Fermi level $E_{F}$ and the mobility edge at the
interface, as indicated in Fig.~\ref{bandalt}. $E'_{F}$, as defined
in Eq.~\ref{energy}, is the quasi-Fermi level at the
insulator-semiconductor interface. By substituting
$dV_{0}=-dE_{a}/e$ in Eq.~\ref{holes} and Eq.~\ref{DOS}, we finally
have the DOS
\begin{equation} \label{DOSfinal}
N(E)\approx\frac{d}{dE_{a}}\left[\frac{\epsilon_{0}\epsilon_{i}^{2}}{\epsilon_{s}l^{2}}U_{g}\left(\frac{dE_{a}}{dU_{g}}\right)^{-1}\right]
\end{equation}
as a function of the energy $E=E_{V}-E'_{F}\approx E_{a}(V_{g})$
relative to the mobility edge.

\subsection{Fraction of free holes and band mobility} \label{demo3}

The fraction of free holes $P_{free}/P_{total}$ is of crucial
importance since it is proportional to the effective field-effect
mobility as described by Eq.~\ref{allgmu}. It can readily be
extracted from temperature dependent measurements. From
Eq.~\ref{pfreeapprox} and Eq.~\ref{energy} and with the total
surface hole density
\begin{equation} \label{ptotal}
P_{total}=C_{i}U_{g}/e,
\end{equation}
we eventually have
\begin{equation} \label{rat}
\frac{P_{free}}{P_{total}}=\frac{L_{a}e}{C_{i}U_{g}}\frac{l}{2l-1}N_{V}\exp\left(-\frac{E_{a}}{kT}\right).
\end{equation}
Finally, from Eq.~\ref{sigmanorm} we see that the band mobility
$\mu_{0}$ can be estimated with
\begin{equation} \label{bandband}
\mu_{0}=\sigma/(eP_{free}).
\end{equation}

\section{Results}

\subsection{Extraction method and the influence of the
contact resistance} \label{demo2}

In this section, we demonstrate the extraction of the DOS and the
hole densities from a set of gated four-terminal measurements.
Moreover, we analyze the influence of the contact resistance on
these functions.

\begin{figure}
\includegraphics[width=0.90\linewidth]{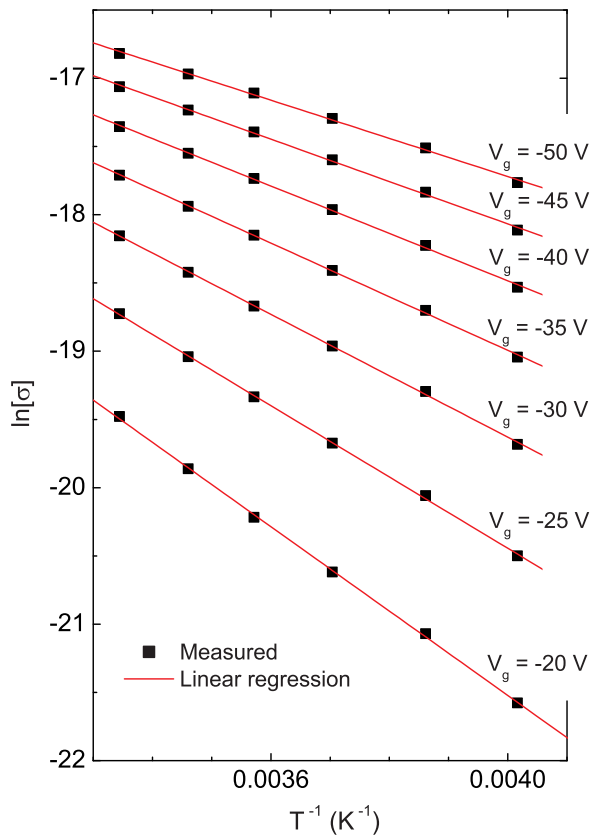}
\caption{\label{arrhenius} (Color online) Arrhenius plots of the
four-terminal conductivity at various gate voltages $V_{g}$. The
activation energy $E_{a}(V_{g})$ was derived from the slope of the
linear regression lines.}
\end{figure}

\begin{figure}
\includegraphics[width=0.90\linewidth]{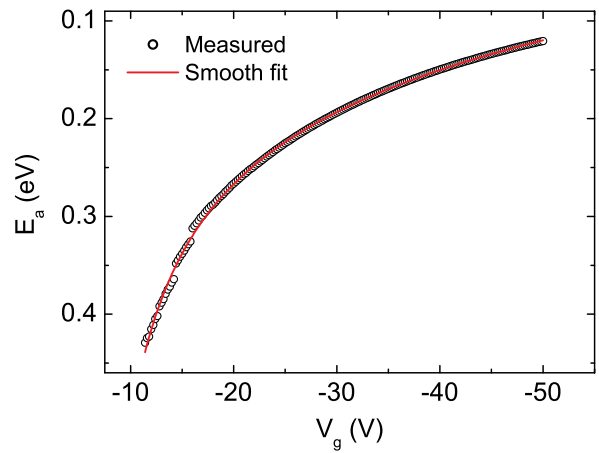}
\caption{\label{activation} (Color online) Activation energy
$E_{a}(V_{g})$ as determined with linear regressions according to
Eq.~\ref{sigmaacti} and as verified with Arrhenius plots
(Fig.~\ref{arrhenius}). The graph also shows a smooth fit of the
activation energy (red/gray line).}
\end{figure}

In a first step, we derived the activation energy $E_{a}(V_{g})$ of
the four-terminal conductivity as a function of the gate voltage,
according to Eq.~\ref{sigmaacti}. Fig.~\ref{arrhenius} shows
Arrhenius plots and the corresponding linear regression lines. We
found that only currents equal to or above $\approx1$\,nA are usable
in the sense that the corresponding four-terminal conductivity
follows a straight line in an Arrhenius plot. Therefore, we used all
currents above 1\,nA for the extraction of the activation energy. At
low gate voltages, only the measurements at the highest temperatures
were considered as a consequence of the 1\,nA limit. The final
result is shown in Fig.~\ref{activation}. The activation energy
$E_{a}(V_{g})$ was then represented by a smooth fit (red/gray line
in Fig.~\ref{activation}) in order to suppress the noise in the
data.

Finally, the DOS was obtained with Eq.~\ref{DOSfinal} and is plotted
as a function of the energetic distance to the mobility edge
$E_{a}(V_{g})\approx E_{V}-E'_{F}$. For the calculations, the
dielectric constant of pentacene was assumed to be $\epsilon_{s}=3$.
In order to determine $U_{g}=|V_{g}-V_{FB}|$ in Eq.~\ref{DOSfinal},
the flatband voltage $V_{FB}$ was assumed to be equal to the device
onset voltage at room temperature. The onset voltage is the gate
voltage where the drain current sharply rises when plotted on a
logarithmic scale, i.e. where the drain current becomes measurable.
The flatband voltage marks the onset of the accumulation regime and
a small difference between the flatband voltage and the onset
voltage may thus exist. A scheme to extract the flatband voltage was
developed for amorphous silicon-based transistors and this scheme
involves the temperature-dependence of the device
off-current.\cite{WeisfieldRL1981} The scheme can, however, not be
applied to our devices because the off-currents are due to
experimental limitations and are not related to the conductivity of
the pentacene thin film.

\begin{figure}
\includegraphics[width=0.90\linewidth]{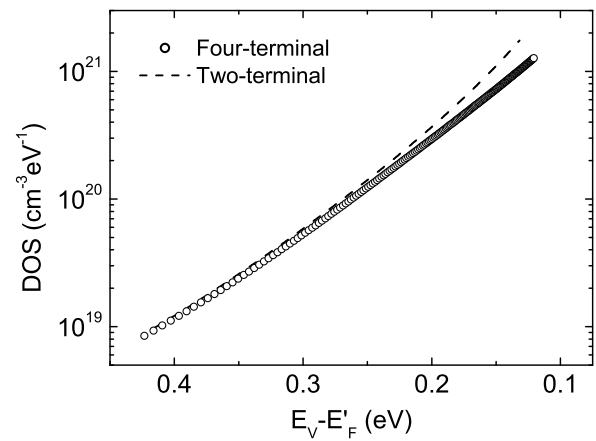}
\caption{\label{DOScomp} Density of traps as a function of energy
relative to the mobility edge (circles). The dashed line was
extracted from the same set of temperature dependent measurements,
but the contact resistance was neglected. The contact resistance can
lead to significant errors in the density of states function,
particularly closer to the mobility edge.}
\end{figure}

Fig.~\ref{DOScomp} (circles) shows the DOS as derived from the
activation energy in Fig.~\ref{activation}. The procedure was also
applied to the same data without correcting for the contact
resistance, i.e. to the two-terminal conductivity. The dashed line
in Fig.~\ref{DOScomp} is the result, highlighting the necessity to
correct for the contacts. Even for long channel devices
($L=450$\,$\mu$m), neglecting the contact resistance leads to
significant errors in the shape and magnitude of the DOS, even more
so closer to the mobility edge.

\begin{figure}
\includegraphics[width=1.00\linewidth]{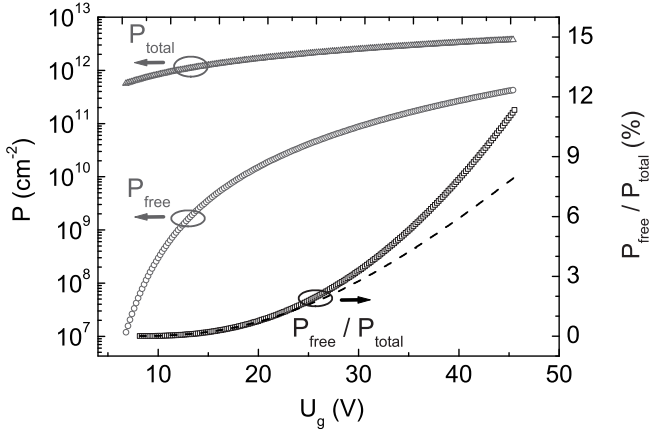}
\caption{\label{ratiometh} Total hole density
$P_{total}=C_{i}U_{g}/e$, free hole density $P_{free}$ and fraction
of free holes $P_{free}/P_{total}$ at room temperature as derived
from the temperature dependence of the four-terminal conductivity. A
significant fraction of the total gate-induced charge is trapped
even at high gate voltages. The dashed line is the ratio
$P_{free}/P_{total}$ if the contact resistance is neglected.}
\end{figure}

The free hole density, the total hole density and the fraction of
free holes was obtained with Eq.~\ref{pfreeapprox}, Eq.~\ref{ptotal}
and Eq.~\ref{rat}, respectively. We have assumed that the effective
density of extended states $N_{V}$ is equal to the density of the
pentacene molecules, i.e. $N_{V}=3\times10^{21}$\,cm$^{-3}$ and
Fig.~\ref{ratiometh} is the result. For the given sample at high
gate voltages, $\approx10$\,\% of the holes that are induced by the
gate are free, i.e. only this fraction actually contributes to the
drain current.

\subsection{Oxygen-related device degradation}

\begin{figure}
\includegraphics[width=0.90\linewidth]{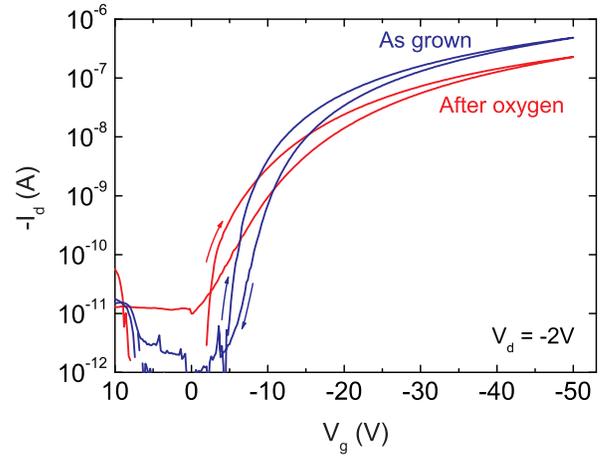}
\caption{\label{compcurrents} (Color online) Linear regime transfer
characteristic of a pentacene thin-film transistor measured as grown
(blue/gray line) and after oxidation (red/gray line). The graph
shows the forward and the reverse sweeps in both cases. The
characteristic oxygen-related degradation effects are a decrease in
subthreshold performance, a decrease in on-current and a shift of
the transfer characteristic to more positive voltages. The current
hysteresis is essentially unaffected.}
\end{figure}

\begin{figure}
\includegraphics[width=0.90\linewidth]{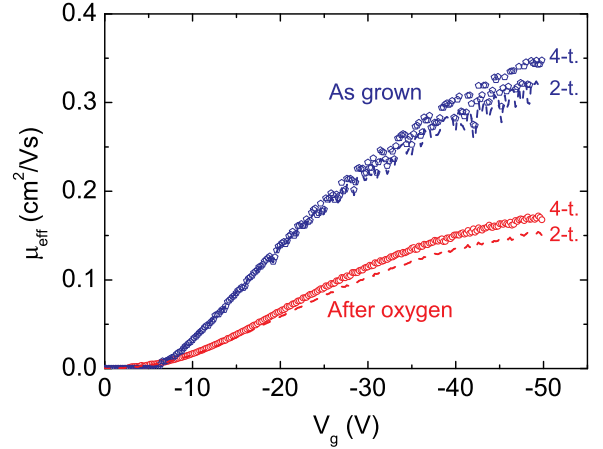}
\caption{\label{compmob} (Color online) Four-terminal effective
mobility (4-t.) from an as grown sample (blue/gray pentagons) and
after the oxygen exposure (red/gray circles). At $V_{g}=-50$\,V for
example, the contact-corrected field-effect mobility decreases from
$\mu_{eff}=0.35$\,cm$^{2}$/Vs to 0.17\,cm$^{2}$/Vs, i.e. it is
reduced by a factor of 2.1. The dashed lines represent the
respective two-terminal mobility (2-t.) where the contact resistance
is neglected.}
\end{figure}

This study correlates the oxygen-related degradation of the
transistor characteristics with the change of the fundamental
transport parameters. We begin by presenting the characteristic
effects of oxygen on the pentacene transistor characteristics. The
blue (gray) line in Fig.~\ref{compcurrents} is a transfer
characteristic measured as grown (after a high vacuum storage time
of approximately 3 weeks). The sample was then exposed to 1\,bar of
oxygen for 19\,h and, additionally, to 1\,bar of oxygen and white
light for 4\,h. The red (gray) curve in Fig.~\ref{compcurrents} is a
measurement of the same device after the oxidation process and after
an evacuation time of 22\,h at a base pressure of the order of
$10^{-8}$\,mbar. Fig.~\ref{compcurrents} contains the forward and
the reverse sweeps in both cases. Fig.~\ref{compmob} shows the
corresponding four-terminal mobilities and, for comparison, the
respective two-terminal mobilities. The degradation effects are the
following: a significant degradation of the subthreshold
performance, a decrease in on-current, a decrease in effective
mobility and a shift of the transfer characteristic towards more
positive voltages. Also the contact resistance is increased after
the oxygen exposure. At room temperature and $V_{g}=-50$\,V it
increases from the as grown value of
$R_{contact}W=2.8\times10^{4}$\,$\Omega$cm to
$9.2\times10^{4}$\,$\Omega$cm after the oxygen exposure, i.e. it
increases by a factor of 3.3. The increase in contact resistance may
be due to the fact that the contact resistance is dominated by the
film resistance. In a top-contact device, the holes must pass from
the electrodes through the pentacene film to the channel at the
insulator-semiconductor interface.\cite{PesaventoPV2006,
KalbWL20072} Importantly, before and after the oxygen exposure we
have a device that is not limited by the contact resistance. At
$V_{g}=-50$\,V for example, the contact resistance $R_{contact}$ is
$\approx14$ times smaller than the channel resistance $R_{channel}$
prior to the oxygen exposure and $\approx9$ times smaller than the
channel resistance after the oxygen exposure.

The degradation effects can be observed when the transistor is
subjected to oxygen in the dark. It is, however, much accelerated
when the oxygen exposure is carried out in the presence of light,
i.e. in the presence of activated oxygen and oxygen radicals.

The degradation of the subthreshold performance and the field-effect
mobility is due to oxygen-related defects that cause electrically
active trap states within the mobility gap. The shift of the
transfer characteristic is due to a change of the flatband voltage.
It is known that oxygen can cause changes of the flatband voltage in
an organic semiconductor device.\cite{MeijerEJ2004, WangA2006}

\subsection{Oxygen-related traps}

\begin{figure*}
\includegraphics[width=1.0\linewidth]{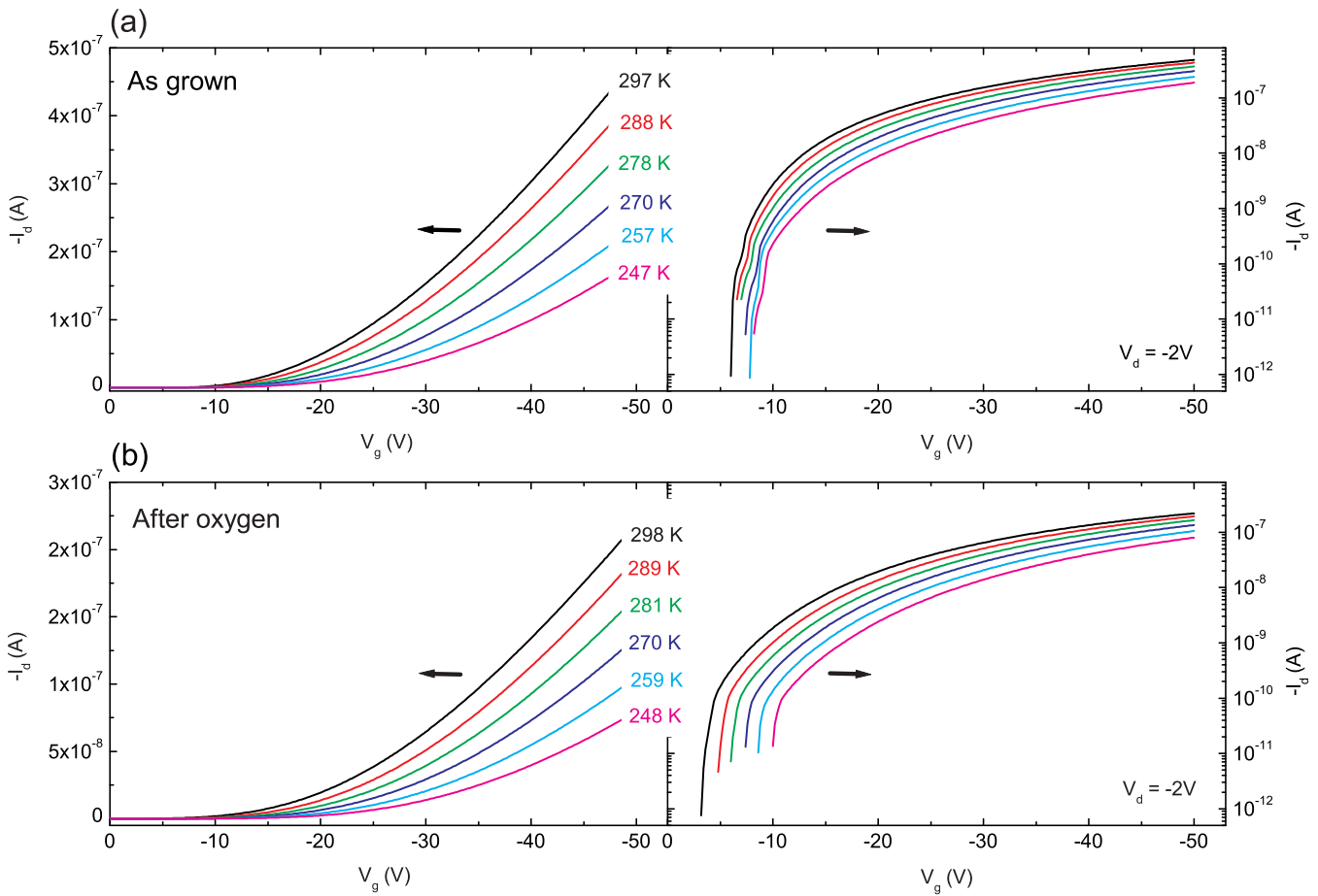}
\caption{\label{together}(Color online) (a) Linear regime transfer
characteristics at various temperatures from an as grown sample,
i.e. prior to oxygen exposure. (b) Transfer characteristics at
similar temperatures after exposing the sample to 1\,bar of oxygen
for 23\,h (19\,h in the dark and 4\,h in white light). The
temperature dependence of the drain current in the subthreshold
regime is much more pronounced after the oxygen exposure.}
\end{figure*}

We now turn to the determination of the trap densities prior to and
after the oxygen exposure. For the temperature-dependent gated
four-terminal measurements, we used a low cooling rate
($0.2-0.25$\,$^{\circ}$/min) in order not to damage the sample. In
Fig.~\ref{together} we show the temperature dependence of the
transfer characteristics prior to and after the oxygen exposure. The
temperature dependent measurements in Fig.~\ref{together} are from
the same sample as the measurements in Fig.~\ref{compcurrents} and
Fig.~\ref{compmob} and were carried out shortly after these
measurements. The main difference after the oxygen exposure is that
the temperature dependence in the subthreshold regime (drain
currents on a logarithmic scale) is much more pronounced.

\begin{figure}
\includegraphics[width=0.90\linewidth]{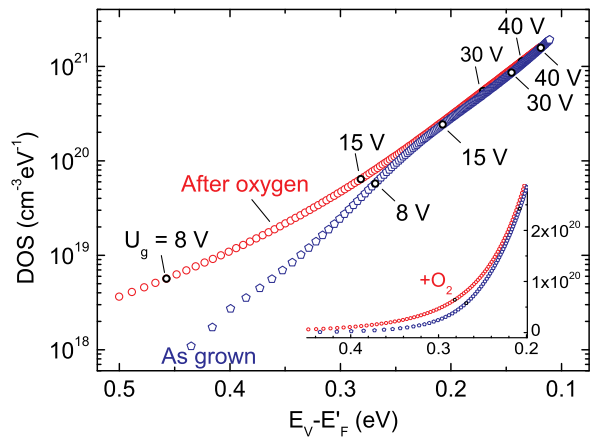}
\caption{\label{DOSgraph} (Color online) Main panel: DOS as a
function of energy relative to the mobility edge on a logarithmic
scale. The blue (gray) pentagons are trap densities measured prior
to the oxygen exposure and the red (gray) circles are trap densities
measured after the oxygen process. The oxidation of the pentacene
film leads to a significant increase in traps that are somewhat
deeper in energy. The corresponding gate voltage $U_{g}$ above
flatband is indicated. The inset shows the deeper traps on a linear
scale.}
\end{figure}

The DOS was extracted for both sets of measurements as described in
Secs.~\ref{demo1} and \ref{demo2}. The main panel of
Fig.~\ref{DOSgraph} shows the final result on a logarithmic scale.
The difference between two adjacent data points in
Fig.~\ref{DOSgraph} corresponds to a change of 0.2\,V in the gate
voltage. Some gate voltages $U_{g}$ are indicated in the graph. The
spacing between the data points decreases as the gate voltage is
increased since at high gate voltages it is increasingly difficult
to shift the quasi-Fermi level $E'_{F}$ towards the mobility edge
due to the increased trap density.

We keep in mind that even in an ideal (trap-free) MIS structure, the
interface potential increases with gate voltage more rapidly at low
gate voltages, than at high gate voltages. This is a screening
effect. The screening depends on the total charge in the device and
it increases with gate voltage.

\begin{figure}
\includegraphics[width=0.90\linewidth]{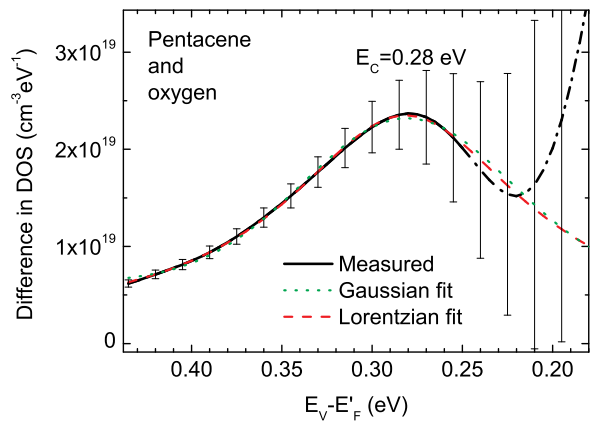}
\caption{\label{peak} (Color online) Difference of the DOS prior to
and after the oxygen exposure (black line). A relative error of
5\,\% is assumed for the determination of a trap DOS and this is
indicated by the error bars. The oxygen exposure leads to a broad
peak of trap states. A Gaussian fit for energies $\geq0.25$\,eV
(dotted green/gray line) gives good agreement with the measured
curve and and so does a Lorentzian fit (dashed red/gray line). The
peak is centered at $E_{C}=0.28$\,eV and the width and height of the
peak are respectively 0.16\,eV and
$2.2\times10^{19}$\,cm$^{-3}$eV$^{-1}$. At energies $\leq0.25$\,eV
the difference in the DOS is comparable to or smaller than the
estimated error and the DOS is essentially unaffected by the oxygen
exposure.}
\end{figure}

The oxygen exposure leads to a significant increase in the density
of traps that are somewhat deeper in energy (Fig.~\ref{DOSgraph}).
The inset in Fig.~\ref{DOSgraph} shows the deeper traps on a linear
scale. In Fig.~\ref{peak} we show the difference of the trap
densities prior to and after the oxygen exposure on a linear scale
(black line). We assume that our method allows for a determination
of the DOS to an accuracy of $5$\,\% and this is indicated by the
error bars in Fig.~\ref{peak}. At energies $\leq0.25$\,eV from the
mobility edge the difference in the DOS is comparable to or smaller
than the estimated error and so for energies $\leq0.25$\,eV, the DOS
is essentially unaffected by the oxygen exposure. At larger
energies, however, the oxygen exposure leads to a broad peak of trap
states. The dotted green (gray) line in Fig.~\ref{peak} is a
Gaussian fit of the experimental data for energies $\geq0.25$\,eV
and the dashed red (gray) line is a Lorentzian fit. In both cases
good agreement is achieved. The Lorentzian fitting function is
centered at $E_{C}=0.28$\,eV and has a width of 0.16\,eV and a
height of $2.2\times10^{19}$\,cm$^{-3}$eV$^{-1}$. The area under the
peak gives a volume trap density of
$\approx4\times10^{18}$\,cm$^{-3}$. With a density of the pentacene
molecules of $3\times10^{21}$\,cm$^{-3}$ this gives an
oxygen-related impurity concentration of $\approx0.1$\,\% provided
that each impurity results in one trap.

The DOS close to the mobility edge is well described by a single
exponential function. A fit for energies $\leq0.22$\,eV gives
essentially identical characteristic slopes prior to and after the
oxygen exposure, i.e. respectively $kT_{0}=47$\,meV and
$kT_{0}=48$\,meV. These values are in good agreement with
characteristic slopes from pentacene-based field-effect transistors
in the literature. A characteristic slope of $kT_{0}=40$\,meV is
reported as determined by simulating the measured transfer
characteristics of pentacene thin-film
transistors.\cite{VoelkelAR2002} Characteristic slopes of
$kT_{0}=32-37$\,meV were derived from pentacene thin-film
transistors with another device simulation
program.\cite{OberhoffD2007, PernstichKP2008} Yet another program
gives a slope of $kT_{0}=100$\,meV for pentacene thin-film
transistors.\cite{ScheinertS2007} However, the band mobility
$\mu_{0}$ needs to be fixed for the simulations and depending on the
choice of the band mobility, slopes of up to 400\,meV are also used
in Ref.~\onlinecite{ScheinertS2007}. With the initial scheme by
Gr\"unewald et al. and from similar pentacene thin-film transistors
as in the present study we have extracted characteristic slopes of
$kT_{0}=32$\,meV shortly after the evaporation of the pentacene and
of $kT_{0}=37$\,meV in the aged thin film with reduced trap
density.\cite{KalbWL20072} For a pentacene single-crystal device, a
characteristic slope of $kT_{0}=109$\,meV is
reported.\cite{LangDV2004}

\subsection{Trap induced changes in the free hole density}

\begin{figure}
\includegraphics[width=1.00\linewidth]{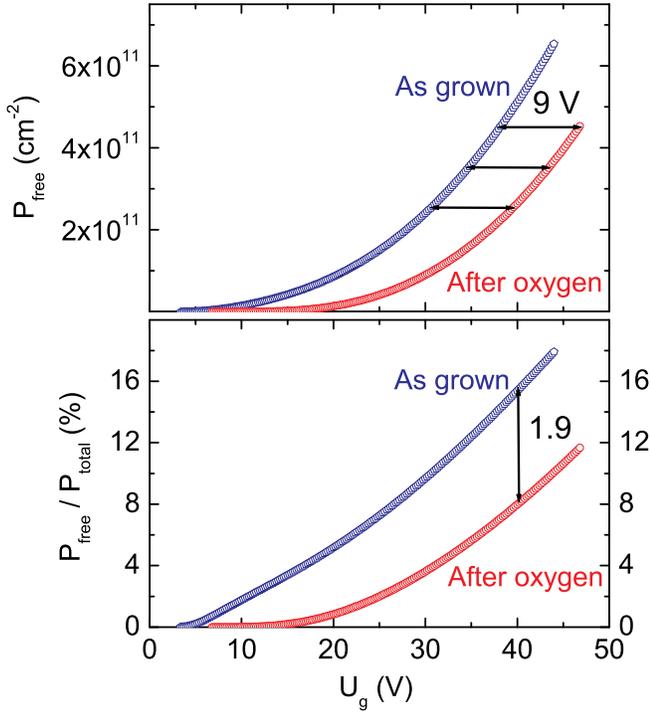}
\caption{\label{ratiocomp} (Color online) Upper panel: free hole
density $P_{free}$ prior to and after the oxygen exposure as derived
from the temperature-dependent gated four-terminal measurements.
After the oxygen exposure the curve is shifted by 9\,V. The
magnitude of the shift is closely linked to density of the
additional traps. Additional trapped holes with a density of
$5\times10^{18}$\,cm$^{-3}$ can be estimated from the shift which is
highly consistent with the trap density estimated from an
integration of the peak in Fig~\ref{peak}
($4\times10^{18}$\,cm$^{-3}$). Lower panel: corresponding fraction
of free holes functions $P_{free}/P_{total}$ prior to and after the
oxygen exposure. At a given gate voltage $U_{g}$, the fraction of
free holes is significantly reduced after the oxygen exposure. At
$U_{g}=40$\,V for example, the fraction of free holes drops from
15\,\% to 8\,\%, i.e. it is reduced by a factor of 1.9.}
\end{figure}

The upper panel in Fig.~\ref{ratiocomp} shows the free surface hole
density (Eq.~\ref{pfreeapprox}) prior to and after the oxygen
exposure from the two sets of temperature dependent measurements in
Fig.~\ref{together}. The parameter $l=T_{0}/T$ was calculated with
the characteristic slopes mentioned above: for room temperature
$l=1.9$ ($l/(2l-1)=0.68$) prior to and after the oxygen exposure. At
sufficiently high gate voltages, the free hole density as a function
of gate voltage is shifted by 9\,V towards higher gate voltages as a
consequence of the oxygen exposure.

The corresponding fractions of the free holes $P_{free}/P_{total}$
were extracted according to Eq.~\ref{rat} and are shown in the lower
panel of Fig.~\ref{ratiocomp}. The fraction of the free holes
changes significantly due to the oxygen exposure. At $U_{g}=40$\,V
for example, 15\,\% of all the induced holes are free prior to the
oxygen exposure and this fraction drops to 8\,\% after the oxygen
exposure. It is reduced by a factor of 1.9. A fraction of free holes
of $8-15$\,\% at 40\,V is in good agreement with values for
pentacene thin-film transistors found in the literature. In
Ref.~\onlinecite{OberhoffD2007}, a fraction of free holes around
10\,\% is specified for comparable total gate-induced charge
densities.

\subsection{Stability of the oxygen-related defects}

\begin{figure}
\includegraphics[width=0.90\linewidth]{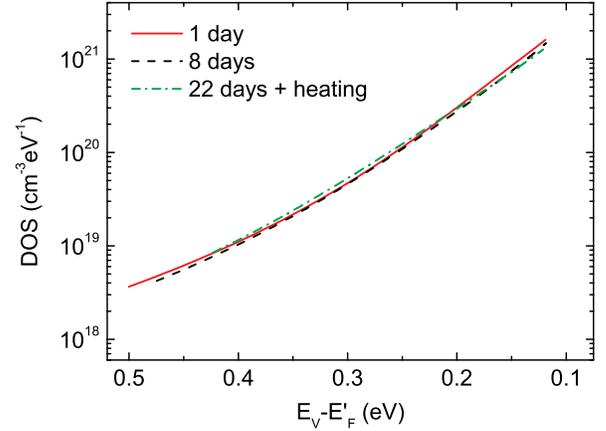}
\caption{\label{dosheating} (Color online) Trap densities after
oxygen exposure. The re-evacuation time after the oxygen exposure is
1\,day (full red/gray line), 8\,days (dashed black line) and
22\,days (dash-dotted green/gray line). The oxygen-related traps are
very stable, i.e. the DOS functions coincide. Prior to the last
characterization after 22\,days, the sample was slowly heated to
temperatures up to 70$^{\circ}$\,C.}
\end{figure}

The DOS after the oxygen exposure was measured after a re-evacuation
time of $\approx22$\,h. In order to elucidate the stability of the
oxygen-related traps, the sample was kept in the prober station at
10$^{-8}$\,mbar for an additional 7\,days period. After that time,
temperature dependent gated four-terminal measurements were carried
out and the DOS was extracted. After these measurements, the sample
was kept at 10$^{-8}$\,mbar for another 10 days. The pentacene films
were then slowly heated to 50$^{\circ}$\,C at a rate of
$0.2$\,$^{\circ}$/min with an electrical heating element at the
cryostat. The temperature was held for 2\,h and the sample was then
left to cool down. The same procedure was repeated with a final
temperature of 70$^{\circ}$\,C. Due to the low heating and even
lower cooling rates, the whole process took 3 days and the effective
heating time was very long. Fig.~\ref{dosheating} shows the DOS
after a re-evacuation time of $\approx1$\,day (same as in
Fig.~\ref{DOSgraph}), 8\,days and 22\,days, the latter time
including the heating procedure. The DOS functions are very similar
and we can conclude that the oxygen-related trap states are very
stable.

\section{Discussion}

\subsection{Effect of oxygen on the trap DOS}

The DOS as extracted from the measurements of the as grown sample is
of particular interest since the sample was kept at a pressure of
the order of 10$^{-8}$\,mbar all along. The trap densities are
relatively high ($10^{18}-10^{21}$\,cm$^{-3}$eV$^{-1}$) with a
rather smooth dependence on energy. In the case of the measurements
on the as grown samples, the effect of ambient gases can be
excluded. It should be kept in mind that we used pentacene powder
that was re-crystallized in high vacuum twice. We conclude that the
``amorphouslike'' trap DOS measured with an as grown sample is
mainly due to structural defects within the pentacene. Trap states
on the surface of the gate dielectric, caused by certain chemical
groups for example, may also contribute to the states that are
deeper in energy.

When pentacene is exposed to oxygen, the gas migrates into the
pentacene film and interacts with the pentacene molecules. This
effect is expected to be accelerated if, in the presence of light,
oxygen is activated and its dissociation is aided. We observe
significant and irreversible changes in the transfer characteristics
and in the DOS caused by the oxygen exposure. It should be kept in
mind, however, that several hours of exposure to 1\,bar of oxygen
are necessary in order to observe these changes. Consequently,
pentacene thin-films are not very sensitive towards oxidation.

The oxygen exposure leads to a broad peak of trap states centered at
0.28\,eV, as shown in Fig.~\ref{peak}. This suggests the degradation
mechanism to be dominated by the creation of a specific
oxygen-related defect. The large width of the peak (0.16\,eV) is
thought to result from local structural disorder that modifies the
on-site energy of the oxygen-affected molecules. As a matter of
fact, very similar arguments are used to explain the smooth
distribution of trap states in hydrogenated amorphous silicon. Even
small deviations in the local structure of a defect lead to a
different electronic state.\cite{StreetRA1991}

Theoretical studies predict various types of oxygen-related defects
in pentacene.\cite{NorthrupJE2003, TsetserisL2007} In
Ref.~\onlinecite{NorthrupJE2003} oxygen defects are discussed in
which a H atom of a pentacene molecule is replaced by an oxygen atom
to form a C$_{22}$H$_{13}$O molecule. The oxygen forms a double bond
with the respective C atom and the p$_{\mathrm{z}}$ orbital at this
atom no longer participates in the $\pi$-electron system of the
pentacene molecule. The oxidation at the middle ring is shown to be
energetically most favorable. These oxygen defects are expected to
lead to trap states in the mobility gap.\cite{NorthrupJE2003} In
Ref.~\onlinecite{TsetserisL2007} other oxygen defects are described.
An example is a single oxygen intermolecular bridge where a single
oxygen atom is covalently bound to the carbon atoms on the center
rings of two neighboring pentacene molecules. This defect, for
instance, is calculated to lead to electrically active traps at 0.33
and 0.4\,eV above the valence band maximum.\cite{TsetserisL2007}

\subsection{Influence of oxygen-related traps on the field-effect mobility}

It is immediately plausible that the oxygen-related traps which are
somewhat deeper in energy lead to a degradation of the subthreshold
performance of the thin-film transistors. We do, however, also
observe a significantly decreased field-effect mobility after oxygen
exposure. This can be understood as follows. The deep traps that are
created by the oxidation need to be filled at first and the position
of the quasi-Fermi level lags behind the position of the quasi-Fermi
level before the oxygen exposure. This is indicated in
Fig.~\ref{DOSgraph} by labeling the corresponding gate voltages
$U_{g}$. At the same gate voltage (which is proportional to the
total gate-induced hole density), the quasi-Fermi level is further
away from the mobility edge. The fraction of free holes, however,
depends exponentially on the position of the quasi-Fermi level
(Eq.~\ref{rat}). The field-effect mobility as described by
Eq.~\ref{allgmu} is proportional to $P_{free}/P_{total}$ and so a
reduction in the fraction of free holes readily affects the
field-effect mobility. At $U_{g}=40$\,V, for example, the fraction
of free holes is reduced by a factor of 1.9 after the oxygen
exposure (main panel of Fig.~\ref{ratiocomp}).

In addition, it is quite possible that the mobility $\mu_{0}$ of the
delocalized charge above the mobility edge is changed after the
oxygen exposure. With Eq.~\ref{bandband}, this mobility is estimated
to be $\mu_{0}=1.2$\,cm$^{2}$/Vs prior to the oxygen exposure and
$\mu_{0}=0.95$\,cm$^{2}$/Vs after the oxygen exposure. We have a
reduction by a factor of 1.3 and a change of the ``intrinsic''
charge transport. Conclusively, the major cause for the reduction of
the effective field-effect mobility is occupancy statistics, and a
reduction of the mobility above the mobility edge also plays a role.

The reduction in the mobility $\mu_{0}$ might be explained by a
scattering of charge carriers at the oxygen-related defects. Another
indication that scattering plays an important role in organic
field-effect transistors is the fact that the mobilities $\mu_{0}$
that we extract are lower than the best field-effect mobilities (up
to 5\,cm$^{2}$/Vs) from pentacene thin-film
transistors.\cite{KelleyTW2003} In addition, repeated purification
of pentacene has been shown to lead to very high mobilities in
pentacene single crystals.\cite{JurchescuOD2004} This effect is
attributed to reducing the concentration of the oxidized pentacene
species 6,13-pentacenequinone which degrades the transport
properties by scattering the charge carriers.\cite{JurchescuOD2004}

\subsection{Trapped holes vs. traps}

The upper panel in in Fig.~\ref{ratiocomp} shows that the oxidation
causes a shift of the curve for the free hole density $P_{free}$ by
$\Delta U_{g}=9$\,V. The same free hole density $P_{free}$ is
realized for different total hole densities $P_{total}$. Clearly,
for an identical number of free holes the difference in the number
of total holes must be attributed to a difference in the number of
the trapped holes. Consequently, due to the oxygen-related traps we
have additional holes that are trapped with a density of
$C_{i}\Delta U_{g}/e=7.5\times10^{11}$\,cm$^{-2}$. Except at very
low gate voltages above the flatband voltage, the charge in an
organic field-effect transistor is concentrated at the
insulator-semiconductor interface. As explained above, our
extraction scheme only considers currents above 1\,nA. Therefore it
is reasonable to assume that the holes are trapped in a region at
the insulator-semiconductor interface with a thickness of the order
of one molecular layer ($\approx1.5$\,nm for pentacene). This gives
a volume density of trapped holes of
$\approx5\times10^{18}$\,cm$^{-3}$. By integrating the peak in the
trap DOS we have derived a trap density of
$\approx4\times10^{18}$\,cm$^{-3}$ which is in very good agreement
with the density of the trapped holes.

\subsection{Deep traps and device performance}

This study reveals how an increase in the density of deeper traps
can significantly affect the field-effect mobility. The influence of
deep traps on the device characteristics is of most general concern
because deep traps can have various origins. Trap states due to the
surface of the gate dielectric, for example, are expected to be
electronically deep traps. Modifying the gate dielectric with a
self-assembled monolayer or using a polymeric gate dielectric not
only leads to an improved subthreshold swing but can also result in
improved mobility.\cite{LinYY19972, VeresJ2004} This also holds in
the case of organic single crystal transistors where the
semiconductor is grown separately and growth-related effects can be
excluded.\cite{PodzorovV2003, GoldmannC20062, KalbWL2007} In the
light of the present study, it seems possible that these effects can
be solely understood with transport in extended states above a
mobility edge and a distribution of trap states: a reduced number of
deep traps leads to an increased number of free carriers above the
mobility edge and to a higher mobility of that charge.

\section{Summary and conclusions}

Pentacene-based thin film transistors were characterized without
exposing the samples to ambient air (as grown) and after exposure to
oxygen in combination with white light. The exposure of the
pentacene to the oxidizing agent leads to a degradation of the
subthreshold performance, a decrease in field-effect mobility, a
shift of the flatband voltage and to an increased contact
resistance.

Contact-corrected trap state functions were extracted from
temperature dependent gated four-terminal measurements. We show that
the exposure to oxygen leads to a broad peak of trap states centered
at 0.28\,eV, with a width of 0.16\,eV and a height of the order of
10$^{19}$\,cm$^{-3}$eV$^{-1}$. The emergence of a peak indicates the
process to be dominated by the creation of a specific oxygen-related
defect. The large width of the peak is due to the energetically
different surroundings induced by structural disorder. The oxygen
defects are very stable and are likely to be caused by pentacene
molecules with covalently bound oxygen.

The decrease in field-effect mobility is caused by the
oxygen-related deep traps. These states are filled upon increasing
the gate voltage and the quasi-Fermi level at the interface lags
behind the position it has in as-deposited samples. This leads to a
significantly smaller fraction of free holes. The magnitude of the
shift in the free hole function is highly consistent with the
density of the oxygen-related traps
($\approx4\times10^{18}$\,cm$^{-3}$) as estimated from the
difference in the trap DOS prior to and after the oxygen exposure.
In addition, the oxygen exposure leads to a decrease of the mobility
of the charge carriers above the mobility edge. This latter effect
may be due to scattering of the charge carriers at the oxygen
defects.

The results can be seen from a more general point of view. At first,
the temperature dependent measurements are self-consistent with the
assumption of a mobility edge, thus contributing to an understanding
of charge transport in organic semiconductors. Moreover, they are an
example for the way in which deeper traps can influence the
effective field-effect mobility.

Theoretical studies may help to identify the oxygen defect and
organic synthetic chemistry may soon find a way to tailor organic
semiconductors where the creation of defects by oxidation is
completely inhibited.

\begin{acknowledgments}
The authors thank Kurt Pernstich, David Gundlach and Simon Haas for
support in the early stages of the study and Thomas Mathis  and
Matthias Walser for stimulating discussions.
\end{acknowledgments}


\newpage

\renewcommand{\baselinestretch}{1.5}
\newpage

\end{document}